# An Intensive Pulsed Neutron Source Based On An Electron Ring Accelerator

S.N.Dolya., K.A.Reshetnikova

*Joint Institute for Nuclear Research, Joliot – Curie street 6, Dubna, Russia, 141980*
*E – mail: sndolya@jinr.ru*

**Abstract**

A neutron source is proposed. It is based on a proton accelerator with the energy $E_p = 1.2$ GeV, current $I_p = 0.7$ A, pulse duration $\tau = 3$ μs, repetition rate $F = 60$ Hz, and accelerator length $L = 30$ m.

Protons are accelerated by the field of electron rings. The electron rings are formed from a tube beam by way of modulation and creating rotating motion by crossing the magnetic field cusp. The frequency of modulation $f_1 = 142.8$ MHz. The bunch is accelerated at the frequency $f_3 = 2856$ MHz. The high frequency power required for making the field is $P_1 = 6 - 150$ MW/section, the power transferred to the accelerated beam is $P_2 = 75 - 100$ MW/m.

The bigger radius of the rings ($r_0 = 2$ cm) and radial dimensions of the ring ($a_r = 0.125$ cm) are provided by the external magnetic field $B_0 = 2.4$ T. The longitudinal dimensions ($a_z < 0.22$ cm) are maintained by the wave moving synchronously with the bunches. The number of electrons in each ring $N_e = 3*10^{12}$, the number of protons $N_p = 3*10^{10}$.

Protons are accelerated at the constant energy gain rate 40 MeV/m. The electron rings are accelerated in the waveguide with the field strength $E_z = 1.08$ MV/m. The intensity of the proton beam is $10^{13}$ p/pulse. The average intensity of the neutron flux on a lead target is $I_{an} = 10^{16}$ n/s, the pulsed neutron flux is $I_{pn} = 8*10^{19}$ n/s.

## 1. Introduction

The modern intensive pulsed neutron sources are based on proton linear accelerators with the energy 1 GeV [1]. These are huge expensive installations with the pulsed proton beam $I_p = 40$ mA. In order to increase the pulse intensity, storage rings are used in such facilities. The total length of such installations is several hundreds meters. The reason for this is that the proton has a small ratio $(Z/A)_p = 1$, compared with the electron $(Z/A)_e = 1840$. Therefore, an idea arises to bind the electron and proton together. But it is impossible to simply «glue» electrons to the proton because electrons will interact with each other, repelling each other, which must be compensated. This is possible to realize if the electrons had the shape of a rotating ring. The proposal to accelerate protons by the electron rings was made by Veksler in 1956 [2].

For a typical number of electrons in the ring $N_e = 3*10^{12}$ and the number of protons $N_p = 3*10^{10}$ we find that the charge of one complex ion is 99e, where e is the elementary charge and the mass of this ion is ($M_p$ + 100 m$\gamma_e$), where $M_p \approx 1$ GeV is the proton mass and m$\gamma_e$ = 15 MeV is the electron mass. The electrons in the ring must have relativistic rotation, in this case the Coulomb repulsion is decreased by a factor of $\gamma_e^2$, where $\gamma_e$ is the relativistic electron factor, and it is possible to receive a compact bunch. So, the total mass of this complex ion is M = 2.5 GeV and (Z/A) ≈ 40. Such a light complex ion can be accelerated by RF power at the frequency f=3 GHz, where the klystron has the power 150 MW and gain near $10^6$ [3].

The possibility to accelerate protons by an electron ring has been widely investigated theoretically and experimentally [4,5]. One of the possible schemes of forming the electron ring is to take a short electron bunch of a tube shape and direct it to cross the cusp of the magnetic field [6,7]. After crossing the cusp, the electrons will acquire a rotating motion and compress in the longitudinal direction in the ratio ($V_{z\ before}/V_{z\ after}$). The bigger radius of the ring $r_0$ and smaller radius $a_r$ practically do not change in this case. If a long-pulsed electron beam receives klystron modulation before the cusp, we can obtain a sequence of rotating electron bunches [8,9]. For example, if the klystron pulse duration $\tau = 7$ μs and the period of modulation $T_m = 7$ ns per one bunch, we will have $10^3$ bunches per one long pulse. If we multiply this by the pulse rate $F = 10^3$ Hz, we will have the intensity up to $10^6$ rings/s. Experimentally, after cusp crossing, the velocity $V_{z\ after}= 0.3$ c has been achieved, where c is the velocity of light, so the ratio of compression ($V_{z\ before}/V_{z\ after}$) = 3 has been obtained [10 -12].

Throughout all the experiments, the problem of maintaining the longitudinal size of the rings has not been solved, which has resulted in defocusing of electron rings in the longitudinal direction and acceleration breakup. Below, a scheme where the longitudinal length of the bunch is kept by the field of a slow electromagnetic wave traveling together with the bunch will be considered.

The scheme of the proposed accelerator is shown in Fig 1.

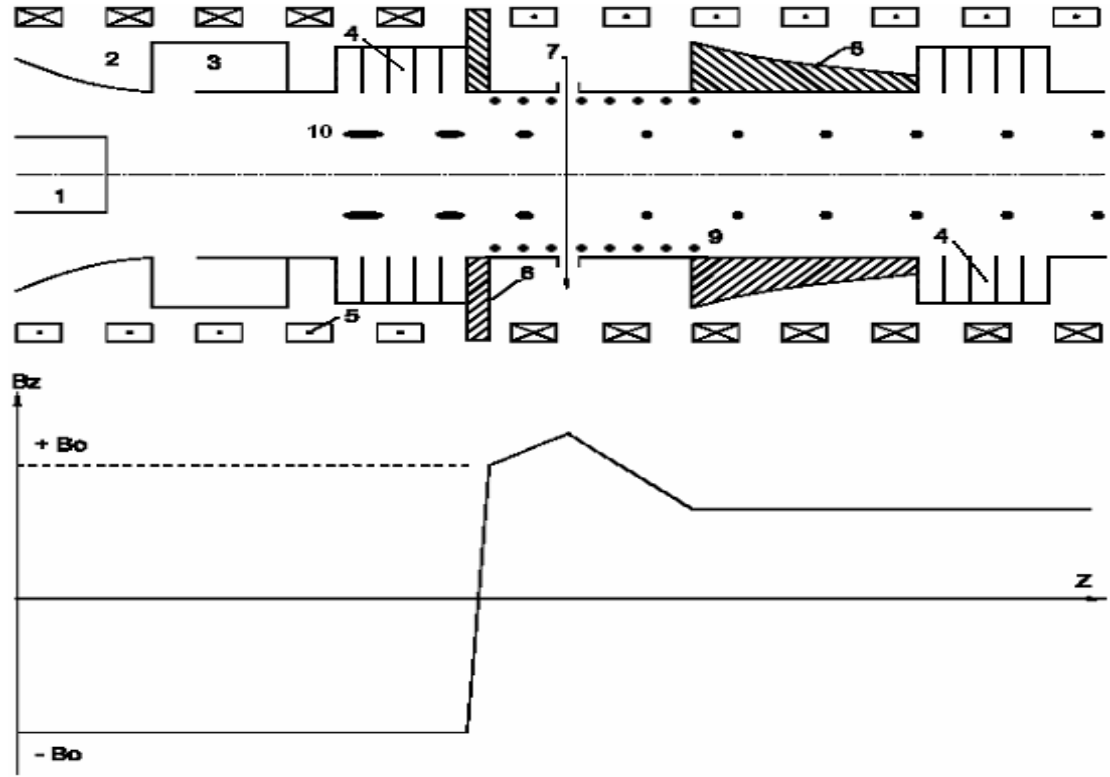

Fig. 1. The scheme of the accelerator.
1 – cathode, 2 – anode, 3 – buncher, 4 – diaphragm-type waveguide, 5 - magnetic coil, 6 – field concentrator, 7 – ultrasonic hydrogen stream, 8 – dielectric waveguide, 9 – spiral waveguide, 10 – electron bunches.

**2. Forming a tube electron beam**

The task of the electron gun is to generate a beam of a tube shape with a current sufficient for electron rings to have the electron number $N_e = 2 - 7*10^{12}$ after klystron bunching. The radius of the tube and radial wall thickness must be equal to the radius of the ring and its small radial thickness. Let us assume that the radius of the beam $r_0 = 2$ cm, wall thickness $2a_r = 0.25$ cm, electron energy $W_e = 200$ keV, electron current $I_e = 140$ A. The typical efficiency of klystron bunching is 50%, so if we bunch this beam at the frequency $f_1 = 142.8$ MHz, we will have $N_e = 3*10^{12}$ electrons in each bunch. A gun with the same parameters, $W_e = 80$ keV, current $I_e = 300$ A, was used [13] for making a sequence of rotating electron rings.

For a ring with the parameters $a=a_r = a_z = 0.125$ cm, we will have the electric field on the surface of the ring:

$$E = eN_e/\pi r_0 a = 70 \text{ MV/m}. \tag{1}$$

## 3. Acceleration of non-rotating bunches

### 3.1 Sub-harmonic bunching

For a resonator with a gap of $h_1 = 1$ cm, the frequency $f_1 = 142.8$ MHz and electric voltage $U_1 = 50$ kV, the distance up to the focus is $l_{f1} = 2.75$ m. At this point another resonator with a gap $h_2 = 1$ cm, the frequency $f_2 = 476$ MHz and electric voltage $U_2 = 40$ kV is placed, then the distance up to the focus $l_{f2} = 35.8$ cm, the semi-width of the bunch in the longitudinal direction being $a_z = 3.45$ cm. Finally, a third resonator with the frequency $f_3 = 2856$ MHz, gap $h_3 = 1$ cm, voltage $U_3 = 10$ kV, length up to the focus $l_{f3} = 22.7$ cm will provide the semi-width of the bunch in the longitudinal direction $a_z = 0.44$ cm.

### 3.2 Acceleration of the non-rotating beam

Acceleration of the beam is conducted at the frequency $f_3 = 2856$ MHz or at the 20$^{th}$ harmonic of the bunching frequency. In this case, 19 separatrices are vacant and only one has a bunch with $N_e = 3*10^{12}$ electrons. The beam will be accelerated in a diaphragm-type waveguide from the initial velocity $\beta = 0.7$ up to the final velocity $\beta = 0.999035$, where $\beta = V/c$. The accelerator length $L = 11$ m.

The RF power can be divided into two parts $P = P_1 + P_2$, where $P_2 = 100$ MW/m provides acceleration of the beam (the pulse current of the beam $I_e = 70$ A, energy rate $\Delta W_e \sim 1.4$ MeV/m) and $P_1 = 6$ MW/section ensures forming of the field. For an accelerator operating with the stored energy, the time between the bunches must be longer than the time of storing the energy in the accelerating section. In this case, the group velocity is small and therefore the lengths of the accelerating sections must be short $L_s = 0.2 - 0.5$ m.

Fig.2 shows how the phase velocity is increased due to change of the diaphragm diameter.

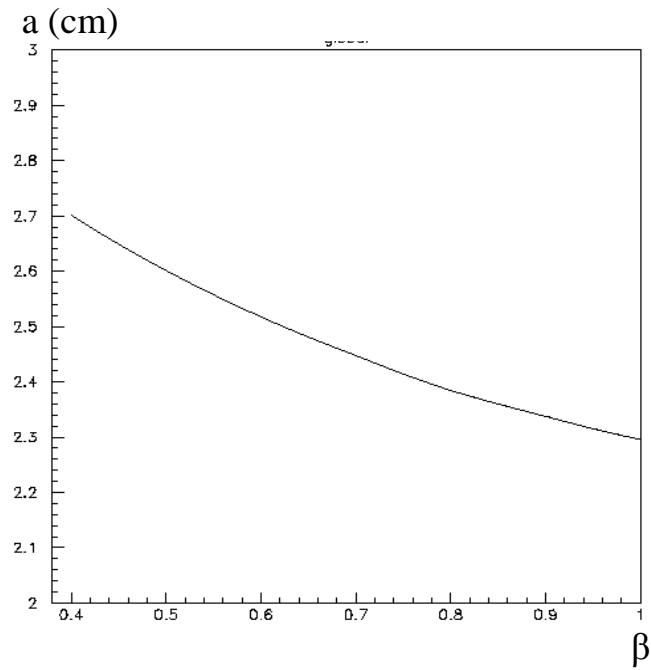

Fig 2. The diameter of the diaphragm required to achieve the needed phase velocity in the diaphragm-type waveguide. The ratio of the outside diameter b to the inside diameter a is b/a = 2.

In Fig. 3, the electric field on the radius of the beam $r_0 = 2$ cm is shown as a function of the phase velocity. In this case, the RF power is $P_2 = 150$ MW/section.

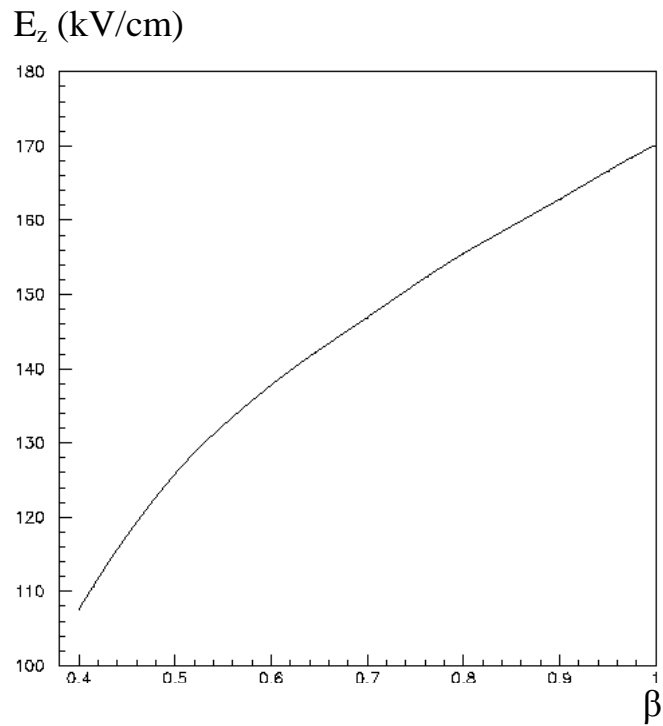

Fig 3. The electric field as a function of the phase velocity.

Fig. 4 shows the dynamics of the semi-width of the bunch in the longitudinal direction $a_z$ as a function of the phase velocity. The amplitude of the accelerating field $E_a = 20$ kV/cm, $\sin\varphi_s = 0.707$.

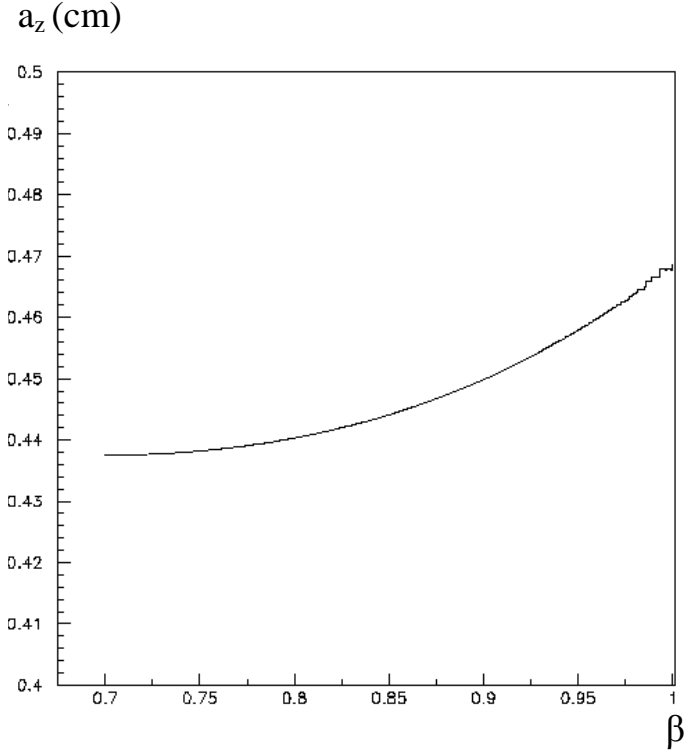

Fig. 4. The semi-width of the bunch in the longitudinal direction $a_z$ as a function of the phase velocity $\beta$.

As is seen, the semi-width of the bunch in the longitudinal direction $a_z$ does not seriously change from the initial value $a_z = 0.44$ cm up to achieving the final velocity $\beta = 0.999035$, corresponding to $\gamma_e = 30$. The Coulomb repulsion is included in the calculation results. The bigger radius of the bunch and the semi-width of the bunch in the longitudinal direction $a_r$ are focused by the external magnetic field $B_0$.

## 4. Cusp crossing

A detailed study of the behavior of cusp-crossing bunches was done in [14], where it was shown that cusp-crossing results in a big spread in longitudinal velocity:

$$\Delta V_z/V_z = (V_{z\,before}/V_{z\,after})^2 (\Delta r/r_0 + \Delta\gamma/\gamma), \qquad (2)$$

where $\Delta r$ is the spread radius at the entrance of electrons into the cusp, $\Delta\gamma$ is the energy spread in the beam. The velocity before the cusp is ultra relativistic $V_{z\,before} \approx c$. It is desirable to have the velocity after the cusp as small as possible, but this would lead to a very big spread in longitudinal velocity $\Delta V_z/V_z$.

In fact, the longitudinal semi-width of the bunch is approximately:

$$a_z \approx \Delta V_z / \Omega_{foc}, \qquad (3)$$

where $\Omega_{foc}$ is the rigidity of focusing. It is not sufficient to simply obtain $a_{z\ after}/a_{z\ before} = V_{z\ before}/V_{z\ after}$, but the semi-width is also to be kept along the whole length of the accelerator. The total longitudinal semi-width of the bunch is:

$$a_z = [a'^2_z + (\Delta V_z/\Omega_{foc.})^2]^{1/2}, \qquad (4)$$

where $a'^2_z$ is the semi-width connected with the kinematical compressions of the bunches and $\Delta V_z/\Omega_{foc}$ is the semi-width related to the rigidity of focusing.

The reason for this is that the big spread in longitudinal velocity after cusp-crossing is due to a small difference between two big values:

$$V^2_{z\ after} = V^2_{z\ before} - r^2\omega^2_B, \qquad (5)$$

the ultra relativistic longitudinal velocity of the electrons before cusp-crossing and ultra relativistic rotating velocity of the electrons after cusp-crossing. The rotating velocity is $r\omega_B = reB/mc\gamma$.

**4.1 The method of the radius spread compensation**

The radius spread compensation can be achieved by moving the bunches through a sequence of cylindrical resonators with the parameters: the gap h = 1cm, frequency $f_3$ = 2856 MHz and inside radius of the resonator a = 2.5 cm. The edge field can be defined by the following formula:

$$E_z = E_0 \exp[-(a-r)/h] \qquad (6)$$

This gap provides faster acceleration to the electrons with a bigger radius than to those with a smaller radius. The amplitude of the field $E_0$ = 100 kV/cm and after twenty gaps the following energy spread will be obtained in the bunches:

$$\gamma(r) = \gamma_0 + \gamma_0 * \Delta r/r_0. \qquad (7)$$

This energy spread accurately compensates the difference in the radius of the electrons' entrance into the cusp.

**4.2 The method of the energy spread compensation**

It is not sufficient to have energy distribution (7) in the beam, at each radius the energy spread must be small. When bunches are accelerated in a traveling wave, the Coulomb repulsion works against the wave field, so the energy spread must be obtained at each radius:

$$\Delta\gamma/\gamma < 5*10^{-4}. \tag{8}$$

# 5. The method of focusing the longitudinal semi-width of the bunch after cusp-crossing

A spiral waveguide is proposed for focusing the longitudinal semi-width of the bunch after cusp-crossing. If the longitudinal velocity after cusp-crossing is $\beta_z = 0.3$ and wave field $E_{zs} = 10$ kV/cm at the frequency $f_4 = 285.6$ MHz, the longitudinal phase frequency $\Omega_{foc}$ will be as follows:

$$\Omega_{foc} = 2\pi f_4 \,[(eE_{zs}\lambda/2\pi\beta_z\mathcal{E}_0)\cos\varphi_s]^{1/2}, \tag{9}$$

where $\cos\varphi_s = 1$ so that this wave can only focus the longitudinal semi-width of a rotating electron ring, $\lambda = c/f_4 = 1.05$ m, $\mathcal{E}_0 = mc^2\gamma_\perp = 15$ MeV is the electron energy. Thus, it is possible to find from formula (9) that $\Omega_{foc} = 4*10^8$, from formula (2) that $\Delta V_z = 6*10^7$ cm/s, and then the longitudinal semi-width connected with the velocity spread is obtained:

$$a_{zsp} = (\Delta V_z/\Omega_{foc}) = 0.15 \text{ cm}. \tag{10}$$

The longitudinal semi-width associated with the kinematical compression is

$$a_{z\,after} = a_{z\,before}\,(V_{z\,before}/\,V_{z\,after}) = 0.15 \text{ cm}, \tag{11}$$

and the total semi-width from formula (4) is $a_z = 0.22$ cm.

In [15], the connection between the flux P and field $E_0$ on an axis in a spiral waveguide was defined, which allowed calculating the electric field on the ring radius, Fig. 5.

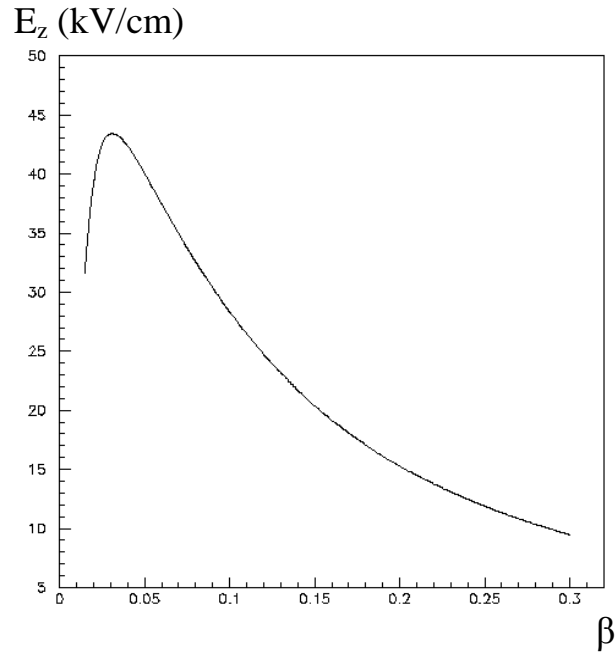

Fig.5. The electric field amplitude $E_{z\,am}$ as a function of the velocity.

At the beginning, the helix pitch distance $h_s$ is very big for creating the phase velocity 0.3 c. The helix radius $a_s$ = 2.5 cm, RF power flux $P_s$ = 5 MW. The flux moves outside the helix and the electric field is small: $E_{z\,am}$ = 10 kV/cm. Then, the phase velocity is to decrease and the electric field at the ring radius grows. After the velocity $\beta < 0.03$, the field "sticks" to the helix so strongly that the field at the ring radius decreases.

The decreasing ring velocity can be achieved due to the motion of a ring up the magnetic "hill", the wave phase velocity is to decrease as the helix pitch distance decreases [15]. Fig.6 shows which helix pitch distance $h_s$ is required in order to receive the needed phase velocity $\beta$. When the ring velocity decreases, the semi-width of the bunch in the longitudinal direction decreases as an adiabatic invariant: $a_z^2 \Omega_{foc}$ = inv and near the phase velocity $\beta = 0.01$ the semi-width of the bunch in the longitudinal direction becomes very small. While crossing the ultra sonic gas stream, electron rings begin to capture protons. After that, electron rings are to be accelerated together with protons up to the velocity $\beta = 0.3$ and the semi-width of the bunch in the longitudinal direction will come back to the initial value.

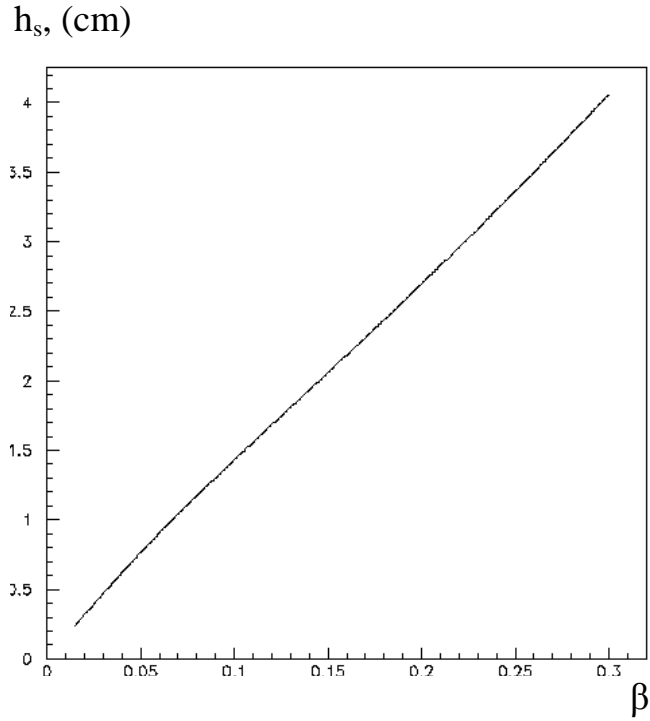

Fig.6. The helix pitch distance $h_s$ required to achieve the needed phase velocity β. The radius of the helix waveguide $a_s = 2.5$ cm.

It is no good decreasing the ring velocity very quickly because in this case the semi-width of the bunch in the longitudinal direction cannot be kept small. In order to slow down a ring as a whole from 0.3 c to 0.01 c, one needs the time $T_f$ to be much longer than the time of the longitudinal oscillation $T_l = 2\pi/\Omega_{foc}$, $T_f \gg T_l$. Now, it is possible to estimate $T_l = 15$ ns, $T_f = 100$ ns and the length of the magnetic field growth $L_g = 4$ m.

## 6. Capture of protons by electron rings and pre-acceleration

For the capture of protons, the electron rings need to be slowed down up to a very small velocity, near 0.01c, where the kinetic energy of the relative motion of protons produced from a neutral gas will be less than the potential energy of protons in the electric field of the rings. The number of protons filling up the electron rings can be estimated by the formula:

$$N_p = N_0 \sigma n_e c \tau, \qquad (12)$$

where τ is the time when the ring crosses an ultra sonic hydrogen stream, $n_e$ is the density of electrons in the ring, σ is the ionization cross–section, $N_0$ is the density of the neutral gas in a stream. In order to calculate $N_p$, let us take the next numerical value: τ = 0.5 ns, c = $3*10^{10}$ cm/s, $n_e = 2.7*10^{12}$ cm$^{-3}$, $N_0 = 6*10^{14}$ cm$^{-3}$, σ = $10^{-18}$ cm$^2$. After that, one can find $N_p = 3*10^{10}$, which is the number of protons required for effective acceleration.

The rings with protons must be pre-accelerated in a decreasing magnetic field at the velocity from 0.015 c up to 0.3 c. The semi-width of the bunch in the longitudinal direction must be kept, in this case, by the field of the helix waveguide, as shown in Fig.7.

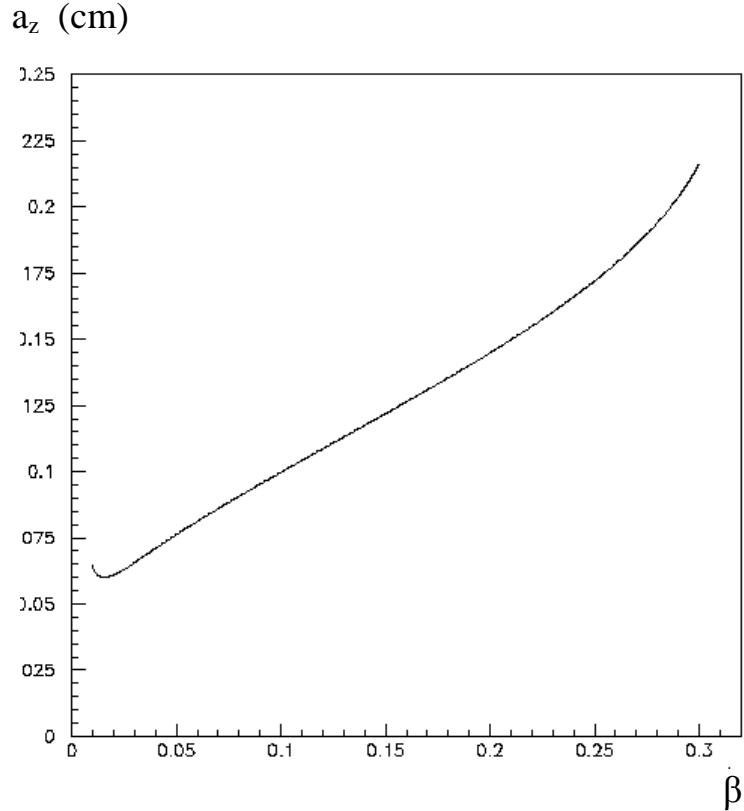

Fig. 7. The semi-width $a_z$ of the electron ring in the longitudinal direction at pre-acceleration at the velocity from 0.015 c up to 0.3 c.

In this case, the gradient of the magnetic field must be small enough to keep the protons inside the electron rings (see below) and, consequently, the length of the pre-acceleration lane $L_p \approx 1$ m.

## 8. Acceleration of the rings in a dielectric waveguide

At the velocity from 0.3 c to 0.5 c, the rings are to be accelerated in a dielectric waveguide [15]. Schematically, the waveguide is shown in Fig.8.

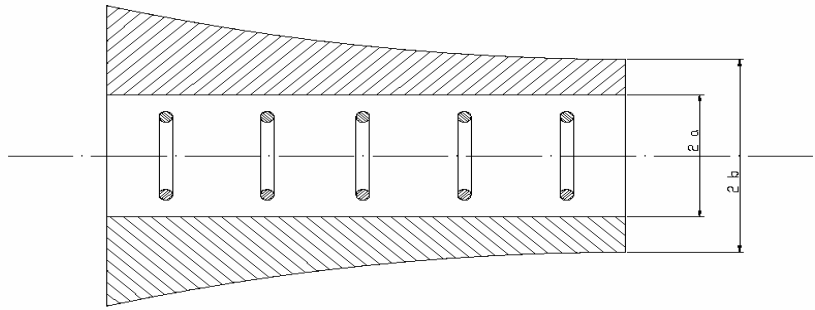

Fig. 8. The scheme of a dielectric waveguide. The region between the outside radius b and the inside radius a is filled with a dielectric with the permeability ε. The region with r <a is free from the dielectric. The phase velocity increases due to the decreasing outside radius b.

A suitable ceramic with the dielectric permeability $\varepsilon' = 30$ and $\operatorname{tg}\delta < 2*10^4$ in the frequency range f < 15 GHz is described in [16]. Next, for the dielectric to be free from charge, it must be covered with a thin metallic film, so thin that it must be transparent for the frequency f = 3 GHz. Now, let us estimate the depth of the skin layer for copper:

$$\delta = c/2\pi\sqrt{\sigma f} = 1.3\mu m. \qquad (13)$$

This problem can be solved if we take the depth 20 – 30 nm.

Fig. 9 shows what the outside radius b of the dielectric waveguide should be like to achieve the phase velocity within the range $0.3 < \beta < 0.5$, with the inside radius of the waveguide a = 2.5 cm.

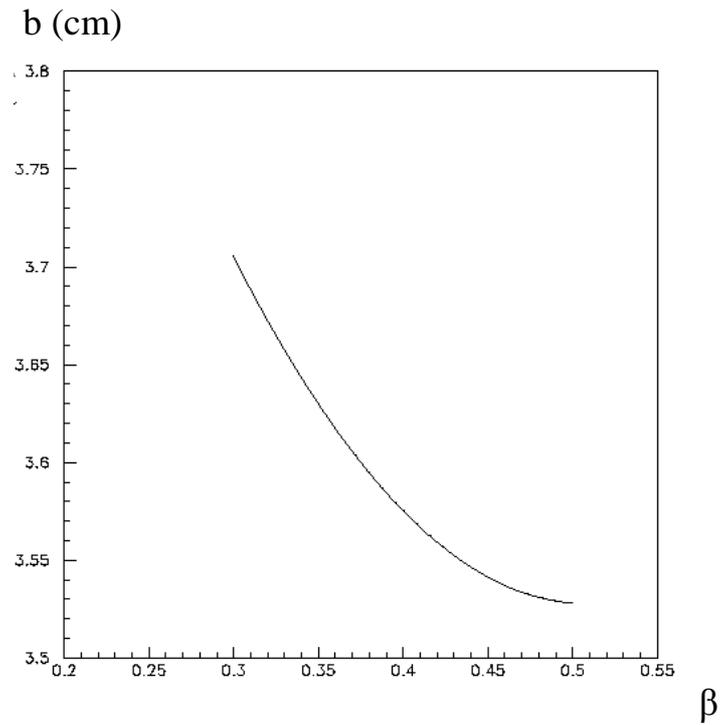

Fig. 9. The outside radius b of the dielectric waveguide required to achieve the needed phase velocity. The dielectric permeability $\varepsilon' = 30$ and the inside radius a=2.5 cm.

Fig. 10 shows the electric field as a function of the phase velocity at the ring radius $r_0 = 2$ cm.

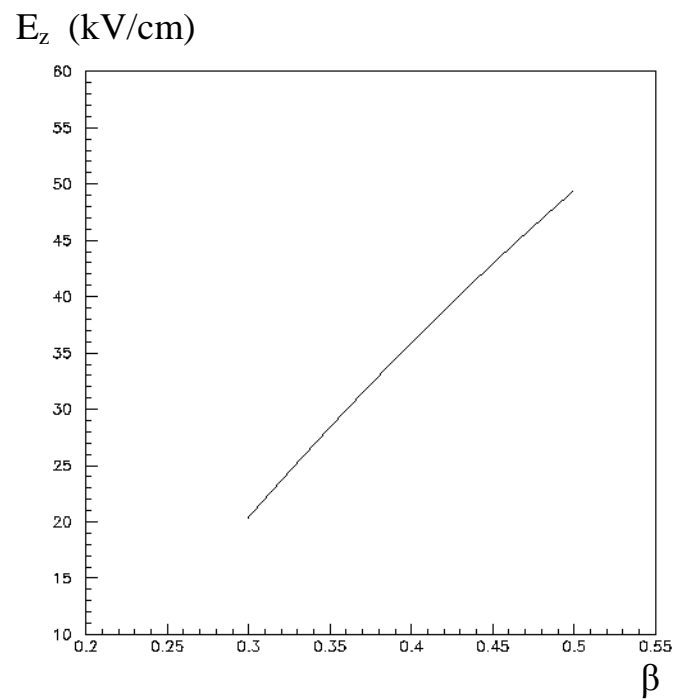

Fig.10. The electric field as a function of the phase velocity at the ring radius $r_0$=2 cm. The RF power is 150 MW/section.

Fig.11 shows the semi-width $a_z$ of the electron ring in the longitudinal direction as a function of the phase velocity.

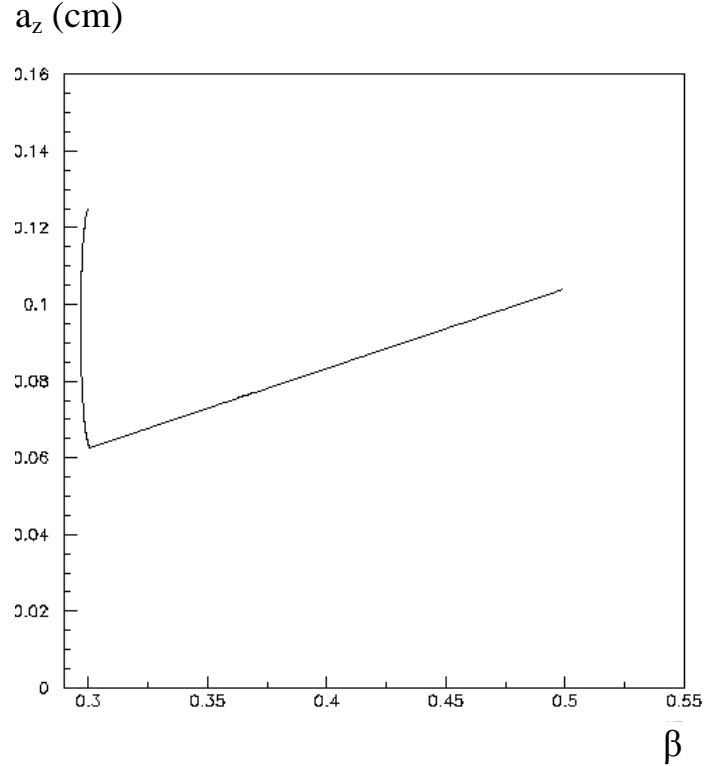

Fig. 11. The semi-width $a_z$ of the electron ring in the longitudinal direction as a function of the phase velocity $\beta$.

## 9. Energy correlations

Let us calculate the electric field of the ring for the following parameters: the number of electrons in the ring $N_e = 3*10^{12}$, bigger radius of the ring $r_0 = 2$ cm, semi-width of the electron ring in the radial direction $a_r = 0.125$ cm, semi-width of the electron ring in the longitudinal direction $a_z < 0.22$ cm:

$$2eN_e a_z/\pi r_0(a^2_r + a^2_z) = 473 \text{ kV/cm.} \qquad (14)$$

We chose the field for the acceleration of protons $E_p = 400$ kV/cm, which is simply an analogue of the synchronic phase $\sin\varphi_s = 400/473 = 0.845$ in classical accelerators. Next, the outside field $E_{out}$ for the acceleration of the electron rings with protons can be found:

$$E_{out} = E_p \, m\gamma_{e\perp}(1+\xi)/M, \qquad (15)$$

where $\gamma_{e\perp}$ is the electron perpendicular relativistic factor, $\gamma_{e\perp} = 28.626$, m is the electron mass, M is the proton mass, $\xi = MN_p/m\gamma_{e\perp}N_e = 0.613$ is the ratio of the total proton mass to the total electron mass. Now, from formula (15) we find the value of the outside field for the acceleration of the electron ring with protons:

$E_{out}$ = 10.8 kV/cm. Correspondingly, $B_r = E_{out}/300 = 36$ Gs, the gradient $B_r = (-r_0/2)dB_z/dz$ is $dB_z/dz = 36$ Gs/cm.

Let us find the power required to accelerate the electron rings with ions. The proton current $I_p = 0.7$ A, the field of the accelerated protons $E_p = 40$ MV/m, hence, the power transmitted to the proton beam $dP_p/dz = 28$ MW/m. At the acceleration length $L_a = 30$ m, the power contained in the proton beam will be $P_p = 840$ MW. The electrons have the final longitudinal relativistic factor identical to protons $\gamma_\parallel = 2.3$, the perpendicular relativistic factor $\gamma_{e\perp} = 28.626$, hence, the total relativistic factor $\gamma = \gamma_\perp \gamma_\parallel = 70$. Now, let us calculate the total electron energy $E_e = 35$ MeV, electron current $I_e = 70$ A, and the total power contained in the electron component $P_e = 35$ MeV $*70$A $= 2.45$ GW. The power transmitted to the electron beam (after crossing the cusp) is 70A*20MeV =1.4 GW. The total power transmitted to the two-component beam after crossing the cusp is 0.84 + 1.4 = 2.24 GW and the rate of power transmission after cusp-crossing is: $dP/dz = 75$ MW/m.

Fig. 12 shows the semi-width $a_z$ of the electron ring in the longitudinal direction as a function of the phase velocity for the acceleration of the ring in a diaphragm-type waveguide.

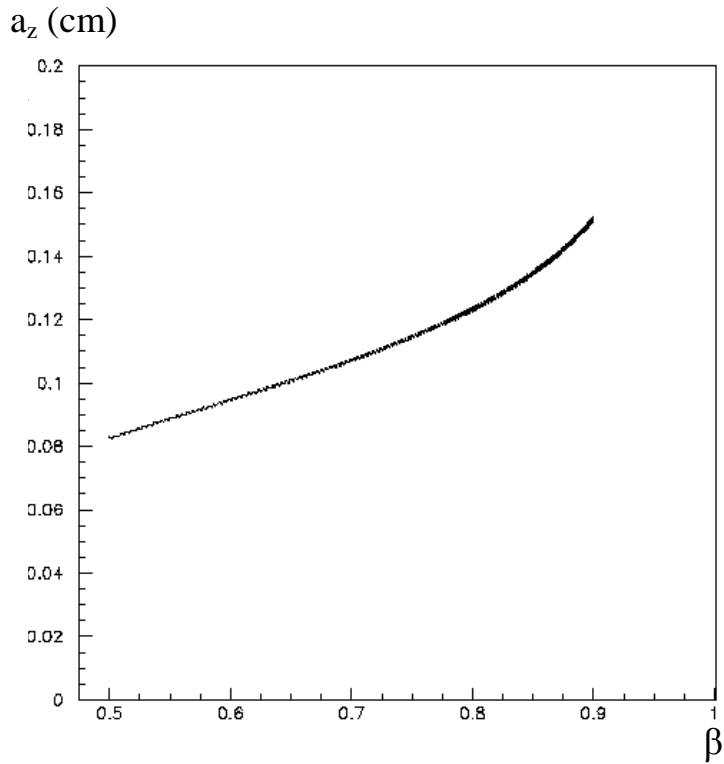

Fig. 12 The semi-width $a_z$ of the electron ring in the longitudinal direction as a function of the phase velocity $\beta$ for the acceleration of the ring in a diaphragm-type waveguide in the range from $\beta = 0.5$ to $\beta = 0.9$.

In the Table, the key parameters of this accelerator are presented.

Table. Key parameters of the accelerator

| | |
|---|---|
| The energy $W_e$ of the electron from the gun, keV | 200 |
| The current $I'_e$ from the gun, A | 140 |
| The radius $r_0$ of the tube beam, cm | 2 |
| The beam wall thickness $2a_r$, cm | 0.25 |
| The frequency $f_1$ of beam modulation, MHz | 142.8 |
| The HF beam current $I_e$, A | 70 |
| The number of electrons in each bunch $N_e$ | $3*10^{12}$ |
| The semi-width $a_z$ of the bunch in the longitudinal direction after bunching, cm | 0.44 |
| The length of the electron accelerator $L_e$, m | 11 |
| The magnetic field T before and after the cusp | 2.4358 |
| The semi-width $a_z$ of the bunch before cusp-crossing, cm | 0.47 |
| The semi-width $a_z$ of the bunch right after cusp-crossing, cm | 0.15 |
| The RF power in a spiral waveguide after cusp-crossing, MW | 5 |
| The frequency of the wave focusing the bunch, MHz | 285.6 |
| The semi-width $a_z$ of the bunch after cusp-crossing, cm | < 0.22 |
| The electric field accelerating the rings with protons $E_z$, kV/cm | 10.8 |
| The electric field accelerating protons $E_z$, MV/m | 40 |
| The length of the proton accelerator, m | 30 |
| The klystron pulse duration, μs | 3 |
| The klystron repetition rate F, Hz | 60 |
| The klystron frequency $f_3$, MHz | 2856 |
| The density of the neutral gas in the ultrasonic stream, $N_0$ | $6*10^{14}$ |
| The final energy of the protons, GeV | 1.2 |
| The number of protons in each ring, $N_p$ | $3*10^{10}$ |
| The pulse current of protons $I_p$, A | 0.7 |

**10. Conclusion**

After the acceleration of protons in the electron rings, the total flux of protons per pulse will be as follows: $P_p = 10^{13}$ protons/pulse.

The electron current is one hundred times greater than that of the proton one but the neutron yield from the electron component is ten times less than the neutron yield from protons. The average intensity of neutrons from the non-multiplying target (lead, mercury) will be $P_n = 10^{16}$ neutrons/s, the pulse neutron intensity $I_n \sim 8*10^{19}$ neutrons/s.

If the average power contained in the beams is increased a hundredfold, for example, the pulse duration is increased up to $\tau = 30$ μs and the repetition rate up to F = 600 Hz, one can achieve the average proton current $\tilde{I}_p = 14$ mA. Such an accelerator (an outside neutron source) can be used as a driver for a subcritical nuclear reactor.

The authors are grateful to L.B.Pikelner for useful discussions.

**References**

1. SNS PARAMETERS LIST: ORNL, SNS-100000000-PL000l-R13, Oak –Ridge, June 2005
2. V.I.Veksler, «Proc. Symp. CERN», 1956, v.1, p.80
3. D.Sprehn, R.M.Phillips, G.Caryotakis, SLAC – PUB - 6677, September 1994
4. V.P.Sarantsev, E.A.Perelstein, «Collective Ion Acceleration with Electron Rings», Moscow, Atomizdat, 1979
5. U.Schumacher, «Collective Ion Acceleration with Electron Rings», Springer, Berlin/Heidelberg, v. 84, 1979
6. M.Reiser, IEEE Tran. Nucl. Sci., v. 18, N3, p. 460, 1971
7. W.W.Destler e.a., IEEE Trans. Nucl. Sci. NS – 22, N3, p. 995, 1975
8. A.G.Bonch – Osmolovsky, S.N.Dolya, Proceedings of $2^{nd}$ International Symposium on Collective Acceleration Methods, JINR, D9-10500, Dubna, 1976, p.87
9. A.G.Bonch-Osmolovsky, S.N.Dolya, Proceedings of $10^{th}$ International Conference on High Energy Charged Particle Accelerators, Serpukhov, 1977, v.1, p.419
10. M.Reiser, IEEE Trans. Nucl. Sci. v. 20, p. 992, 1973
11. C.D.Striffler e.a., IEEE Trans. Nucl. Sci. NS – 26, N3, p. 4234, 1979
12. D.W.Hudgings e.a., Phys. Rev. Letters, v.40, N 12, p.764, 1978
13. S.N.Dolya, A.K.Krasnykh, V.V.Tikhomirov, Meeting on Collective Acceleration Method Problems, Dubna, 1982, D9-82-664, p. 47-49
14. A.G.Bonch-Osmolovsky, S.N.Dolya, JINR, Dubna, 1976, 9-10228
15. A.I.Akhiezer, Ya.B.Fainberg, UFN, v.44, issue 3, p. 322, 1951
16. S.Kawashina e. a., J. Am. Ceram. Soc. V. 66, N6, p. 421, 1983